\documentclass[%
 reprint,
 amsmath,amssymb,
 aps,twocolumn,pra
]{revtex4-1}

\usepackage{graphicx}
\usepackage{hyperref}
\usepackage{multirow}
\usepackage{dsfont}

\newcommand{\eqn}[1]{\begin{eqnarray} #1 \end{eqnarray}}
\newcommand{\tit}[1]{\textit{#1}}
\newcommand{\tbf}[1]{\textbf{#1}}
\newcommand{\trm}[1]{\textrm{#1}}

\newcommand{\tr}[1]{  \textrm{Tr}\left[ #1 \right]  }
\newcommand{\trc}[2]{  \textrm{Tr}_{#1}\left[ #2 \right]  }
\newcommand{\zum}[2]{\displaystyle\sum_{#1}^{#2}}

\begin{document}

\title{Extending the Agent in QBism}

\author{Jacques Pienaar}
\affiliation{QBism group, University of Massachusetts Boston, 100 Morrissey Boulevard, Boston MA 02125, USA.
}

\date{\today}



\begin{abstract}
According to the subjective Bayesian interpretation of quantum mechanics (QBism), the instruments used to measure quantum systems are to be regarded as an extension of the senses of the agent who is using them, and quantum states describe the agent's expectations for what they will experience through these extended senses. How can QBism then account for the fact that (i) instruments must be calibrated before they can be used to `sense' anything; (ii) some instruments are more precise than others; (iii) more precise instruments can lead to discovery of new systems? Furthermore, is the agent `incoherent' if they prefer to use a less precise instrument? Here we provide answers to these questions.
\end{abstract}

\maketitle



\section{Introduction}

QBism is an interpretation of quantum mechanics according to which quantum states quantify an Agent's subjective degrees of belief about their future personal experiences (Events). According to QBism, quantum theory itself is not meant to describe an external world independently of observation; rather, quantum theory is best understood as a normative addition to standard probability theory, that is to say, it is a rule constraining an Agent's subjective probability assignments for the results of measurements on quantum systems. Normative rules are founded upon the principle of coherence: an Agent must strive to hold beliefs that are `Dutch book coherent' (that is, which would not imply a sure loss if the Agent were to make bets according to them), and that are coherent with the world (i.e.\ that accommodate lessons learned from past experiences). 

QBism is thereby able to avoid at least one aspect of the measurement problem, the `problem of definite results', which seeks to explain why a measurement of a quantum system leads to a single result. QBism takes the personal experience of any Agent as fundamental; as such, the occurrence of definite results is a basic postulate that does not need to be derived or explained within the theory. Moreover, by insisting that the relevant definite results are Events that are personal to the Agent who experiences them, QBism also evades the teeth of the dilemmas posed by the EPR `paradox' and Bell's theorem \cite{FDR2014}. 

Despite these advantages, QBism's radical stance on the nature of measurement renders it vulnerable to another line of attack, concerning the status of the measuring apparatus. The question is a simple and natural one: should the physical devices that we scientists employ in order to obtain the results of our measurements be themselves treated as measured systems external to the Agent who uses them, or should they be treated as being on par with the perceptual organs of the Agent, giving them an essentially direct experience of the measured system? Put in concrete terms, when a scientist uses a Stern-Gerlach apparatus to discern the spin of an electron, is the relevant Event that the electron is experienced as having, say, spin `up', or is it that a small dot is experienced to appear on the upper half of the glass screen?

To their credit, QBists have been unequivocal in their answer, which is that the apparatus is to be regarded as essentially part of the Agent. This standpoint represents a key way in which QBism departs from a Copenhagen-like interpretation in the spirit of Bohr. For example, Fuchs writes \cite{Fuchs2017}:

\begin{quote}
    QBism, in contrast, takes the idea of `the instruments of observation as a ... prolongation of the sense organs of the observer' deadly seriously  and runs it to its logical conclusion. This is why QBists opt to say that the \tit{outcome} of a quantum measurement is a \tit{personal experience} for the agent gambling upon it. Whereas Bohr always had his classically describable measuring devices mediating between the registration of a measurement's outcome and the individual agent's experience, for QBism the outcome \tit{just is} the experience.
\end{quote}

Stranger than the QBists' firm adherence to this somewhat unintuitive stance is the fact that nobody has criticised it. Yet it is a ripe target: do QBists really think that a scientist directly experiences the spin of an electron to be `up', in a manner not fundamentally different to experiencing, say, that the orientation of a weather-vane is North? If that is so, then how does the QBist account for the fact that the apparatuses used to experience Events typically require laborious tuning and calibration, and must be used by a skilled and highly trained individual, whereas the orientation of weather-vanes can be checked effortlessly by any sighted person? Furthermore, the QBists would have it that an Agent equipped with an apparatus is to be treated as essentially distinct from an Agent without such equipment. Yet no scientist enters the lab for the first time already proficient with the equipment, and once they are proficient, it seems peculiar to say that they have become an entirely different Agent. What, then, is the connection between the equipped and unequipped Agent? 

Finally, QBism has provided no clarification as to what makes one apparatus superior to another. This means the QBist, to date, has no compelling counter to the conventional narrative that measuring devices of increasing refinement more closely approximate the `true' values of the measured quantities. Since QBism rejects the notion that there is an independently exisiting and objective `true' value of any quantity, it is apparently unable to explain how scientific instruments can progress to ever increasing levels of precision and refinement. If QBism is to meet this challenge, it must provide a criterion for comparing apparatuses and identifying which ones are `more precise and refined' than others, compatible with QBism's tenets. The aim of this work is to fill in this gap by supplying precise answers to all the above questions within the QBist interpretation.

\tit{Outline of the paper:} In the next part, we give a detailed overview of QBism, its key concepts and formal structure. In section \ref{Sec:Extension} we further clarify what is meant by an `extension of the agent' on a QBist interpretation, and propose a formal definition. In Sec.\ \ref{Sec:comparison} we make a connection to the resource theory of quantum measurements and use it to classify and compare the different types of Agent extensions. In particular, we provide a QBist notion of `refinement' that makes no reference to pre-existing `true' values of measured quantities.\\ 

\section{Overview of QBism \label{Sec:Overview}}

In this section, we give a self-contained introduction to QBism, covering its basic premises including its treatment of measurements, coherence, and the Born rule. We will find it necessary to further elaborate on the QBist definition of a `system' and on the notion of the Agent's physical body in more detail than can be found in previous works.

\subsubsection{Event, Agent and World}

In QBism, an element of reality is an experience, which contains as a fundamental internal structure a pairing of an experiencing subject with an experienced object; such experiences are called Events. (In this work, we use capitalization to denote concepts that are personal to the experiencing subject; thus the personal Events are to be distinguished from the impersonal or objective `events' commonly found in physics textbooks.) Take, for instance, the Event of biting into a cookie. The features of the Event may include the taste of raisins, the feeling of surprise and disappointment at not tasting chocolate chips, the feeling of grasping the cookie, and so on. Among other things, these features implicate an Agent (the one who tastes, feels, etc.), who has a body (a tongue to taste, hands to grasp, etc.), and also a personal external World (the thing to be tasted and grasped, i.e.\ the target of the Agent's experiences). The notions of Event, embodied Agent, and World, are all fundamental in QBism. They frame and underpin all subjective experiences, from mundane commonplace experiences like eating cookies, to sophisticated and theory-laden experiences like measuring quantum systems. Even more, the extent to which scientific knowledge is `objective' must be understood as arising from the inter-subjective experiences of many Agents.

Every Event refers to some particular Agent, and so it is both real and personal to the Agent that experiences it. The features of an Event are what is `directly experienced' by the Agent, that is, immediately and without the need of conscious reflection or interpretation. Note that the QBist notion of `direct experience' is intended to be rich enough to capture aspects of perception beyond mere `sense-data'; the Agent's emotions, memories, and expectations all play a role in shaping this experience.

For example, a chess grandmaster looking at the state of a game in progress may be said to directly experience that Black is in a strong position, because this may be clear to them at a glance. We would not say the same of a novice, who would only directly experience the positions of the pieces on the board, and would have to do conscious work to deduce or infer Black's advantage in the game. (Incidentally, there is perhaps a sense in which the grandmaster's training has given them a `mental apparatus' not unlike a physical apparatus.)

Since the Agent and World represent internal aspects of Events, one should be careful not to think of the Agent and the World as being \tit{causes} of the Events, at least not on any standard reading of causality that physicists have used; we will say more about this in Sec.\ \ref{Sec:mmts}. While the meaning of causality in QBism has yet to be fully explicated, this much can be said: the status of an individual Event in QBism is comparable to that of the `big bang' in standard cosmology, in that it has no causes within the theory that describes it, and may be regarded for theoretical purposes as essentially un-caused. In the case of QBism, of course, there is not one but a multitude of such primordial Events that make up the fabric of reality. The Events are, to use Wheeler's words, `elementary acts of observer participancy', which led him to say `yes' when asked `is the big bang here?' (see \cite{Wheeler} and discussion in \cite{FuchsNWB}). 

This might seem to pose a challenge to the idea of scientific explanation, but this can be met provided one is willing to adopt a more flexible notion of what `explanation' means. In QBism, scientific explanation does not seek to identify the causes of the elements of reality; rather, the occurrence of Events is a basic premise from which scientific analysis proceeds. The explanation for a phenomenon, such as the occurrence of an unexpected Event, is to be sought in making it coherent with the Agent's interconnected mesh of beliefs. The Agent seeks to achieve this by adjusting their interpretation of the meaning of the Event and/or revising their existing beliefs. An unexplained event is one that has low probability given the Agent's beliefs and past experience; in short, an occurrence that the Agent continues to find surprising even in hindsight. An explained event is the opposite: one whose occurrence, though it may have surprised the Agent at the time, is attributed a high probability in light of the Agent's current (possibly revised) beliefs. To give more detail than this brief sketch would take us beyond the scope of the present work.

One of the stated goals of QBism is to clearly separate what is subjective and objective in quantum theory. To do this, QBism calls for a radical revision of what these terms mean in physics. Unlike other interpretations, QBism does not regard `real' as being synonymous with `objective'. QBism begins with the premise that reality is intrinsically subjective, and proceeds to ask how objective knowledge comes about in this context. The QBist interpretation of quantum theory starts from the assertion that quantum states represent an Agent's subjective probabilities about subjective Events that may happen to them. The objective content of quantum theory therefore is not localized in states or measurement results, but rather resides in more holistic structural features of the theory that apply equally to all Agents. The identification of objective features of this kind is one of the main ongoing projects of QBism.

\subsubsection{Measurements \label{Sec:mmts}}

Given that Events are infinitely rich and variable in their qualities, it is not feasible to enumerate all possible Events that may occur at the next moment, let alone assign them probabilities. The first step in making sense of the world as a scientist is to quantify the possibilities as belonging to some discrete set. Thus, when I measure the length of a table, I do not consider the innumerable different experiences I might have in measuring it, nor even the innumerable ways that I might perform the measurement -- I only consider the single coarse-grained action of `measuring the table length', and the resulting value in (say) millimeters that I am able to discern with my eye. I judge the table to be between $1000$ and $2000$ mm, so although my possible experiences on measuring the table are innumerable, there are only $1000$ distinct possible Events that will describe my \tit{quantified} experience of `measuring the table length'. More generally, a measurement refers to an Event that is characterized by the following two features:\\
(i) A quantified deliberate action by the Agent (the `measurement action');\\
(ii) A quantified response of the World (the `measurement result').\\
We avoid using the terms `setting' and `outcome' for these categories because common usage would have a `setting' refer to, say, the experimenter's turning a dial to some value on an apparatus, and the associated `outcome' to some uncontrolled or dependent variable, say, the triggering of a photon detector, but this usage does not correspond to our notions of action and result. According to our usage, the `setting' refers to an Event, i.e.\ an action-result \tit{pair}, whose action is turning the dial, and whose result is seeing the dial come to rest at a certain position; the `outcome' refers to another Event whose action is to listen for the clicks of the detector at some time, and whose result is hearing the detector click (or not). The difference between the two is not a matter of type, but merely of the degree to which the Agent judges the measurement result to depend on the their voluntary actions in each case. For instance, in turning a dial with our hand, we generally expect to see it come to rest at the intended value, and so would be inclined to call the Event a `setting'; by contrast, our expectation about whether the detector will click is usually judged to be independent of our voluntary act of listening for the click, and so the Event tends to be classified an `outcome'. 

Furthermore, whereas it is standard to regard `settings' as causes of `outcomes', an action is not interpreted na\"{i}vely as the cause of its associated result, because both arise together in the very moment that the Event occurs to which they pertain. This kind of relation is not causal in a deterministic sense, nor is it causal by the definitions commonly used in statistical causal modeling (e.g.\ that of Pearl \cite{Pearl}). While it is possible to define causal relations of the latter kind \tit{between} measurement Events such as settings and outcomes (see Ref.\ \cite{PIEQBcausal} for a program along these lines), the same cannot be said for the action-result pair, which co-arise together within a single Event. 

We mention that this kind of `co-arising' relation does have precedents in the philosophical literature. According to William James: 
\begin{quote}
Your acquaintance with reality grows literally by buds or drops of perception. Intellectually and on reflection you can divide these into components, but as immediately given, they come totally or not at all. \cite{WJ}
\end{quote}
Similarly, Whitehead spoke of `actual entities' \cite{Whitehead}, Husserl of `intentional experiences', and Merleu-Ponty of `sensations'. All of these ideas reject a formal splitting into `cause' and `effect'. For instance, Merleu-Ponty asserts that
\begin{quote}
in this transaction between the subject of sensation and the sensible it cannot be held that one acts while the other suffers the action, or that one confers significance on the other. \cite{MP}
\end{quote}
This shows that philosophy may provide further insight as to how QBism might interpret relations that are internal to Events, such as the action-result and Agent-World relations.


To quantify an Event, the Agent decides upon a finite, discrete set of possibilities for the measurement action (say, `measure the table length' or `set the oven dial to $250$ degrees') and similarly for each measurement result (say $1000$-$2000$mm, and $100$-$400^{\circ}$, respectively). Explicitly, let $a \in A$ denote the set of possible actions and $x \in X^{a}$ the set of possible results of action $A=a$. The sample space consists of the quantified Events $\{ (X^{a} = x) : \forall x \in X^{a}, \forall a \in A \}$, which we label as ordered pairs $(a,x)$. For each possible action, the Agent assigns probabilities $\{ P_{a}(x) : \forall x \in X^{a} \}$, which quantifies how likely they think the Event $(a,x)$ is to occur. By assumption, the Agent can only be uncertain about the value of the feature $x$ (which belongs to their external World) and not about the value $a$ of their own action. Consequently, the probabilities satisfy
\eqn{
\zum{x \in X^a}{} P_{a}(x) = 1 \, , \qquad \forall a \in A \, .
}

\subsubsection{Coherence}

QBism regards probabilities as a means of quantifying the subjective degrees of belief of an Agent. As such, QBism embraces a subjective Bayesian interpretation of probability theory. On the subjective Bayesian account, there is no such thing as a `right' or `wrong' probability assignment, not even by degrees. Rather, the emphasis is placed on whether or not an Agent's set (or `mesh') of beliefs is internally \tit{coherent}. Coherence is understood as a \tit{relation} among beliefs, so while it does not constrain any individual belief, it constrains an Agent's beliefs as a whole and serves to define an organizing structure for belief that is equally applicable to every Agent. In practice, coherence is defined negatively, by positing the conditions under which it is violated. The strongest notion of coherence is \tit{Dutch book} coherence. This type of coherence is violated if, given the Agent's probability assignments for a set of Events, a `Dutch book' can be constructed from them, which is a set of bets (say, proposals to buy or sell lottery tickets to the Events at certain prices) such that the Agent considers each bet worthwhile when taken individually, but when taken together they imply a sure loss for the Agent. It can be shown that failure to conform to the basic axioms of probability theory (e.g.\ making probability assignments that are negative or don't sum to one) implies the existence of a Dutch book, and so the probability calculus can be justified on the grounds of Dutch book coherence \cite{Jeffreys}.

Dutch book coherence alone is not enough to capture all that is meant by `coherence' in an Agent's beliefs. The most prominent example is the rule of Bayesian updating, which determines the probability $P_1(H)$ an Agent should assign to an Event $H$, in light of some new Event $E$, given that the Agent's beliefs prior to $E$'s occurrence were given by $P_0(H),P_0(E),P_0(E|H)$. The rule of Bayesian updating stipulates that:
\eqn{ \label{eqn:updating}
P_1(H) &:=& P_0(H|E) \, ,
}
where $P_0(H|E) = P_0(E|H) P_0(H) / P_0(E)$. (The latter identity, which relates beliefs at a single time, should not be confused with Bayesian updating just described, which relates the Agent's present beliefs to their prior beliefs at an earlier time). Failing to update one's beliefs according to \eqref{eqn:updating} does not make one vulnerable to a Dutch book, and indeed there can be situations where ignoring the rule \eqref{eqn:updating} makes sense \cite{FS_reflection}. What then could motivate an Agent to impose the constraint \eqref{eqn:updating} on their beliefs? 

It seems that coherence is about more than just striving to avoid a Dutch book: it is also about adhering to certain constraints on one's beliefs that are inspired by perceived regularities in one's experience. For instance, Galileo's equation for the height of a falling body released from rest may be thought of as expressing a normative rule: that the Event of releasing the object from a fixed initial height would constrain an Agent's beliefs about it's height at subsequent times. Unlike Dutch book coherence, which seems largely independent of the particular features of the world we live in, normative rules like Bayesian updating or falling under gravity are more contingent on one's actual experiences. For example, an astronaut would not find reason to adhere to the `normative rule of falling' just described. We therefore propose that such normative additions to probability theory, including the rule of Bayesian updating, should be interpreted as being founded on `World-coherence', as distinct from Dutch book coherence.

\subsubsection{The Born rule}

According to QBism, the difference between quantum and classical theory hinges on the form of a particular normative addition to probability theory that describes how the results of a hypothetical measurement (one that is not actually performed) constrain the Agent's beliefs about an actual measurement.

To illustrate, consider a possible measurement that you could perform on a box of chocolates, say, counting them. Suppose that you are presented with a closed box, for which you assign the prior probability $p(i)$ that you would find $i$ chocolates inside, if you were to open it. Now imagine that you are asked to weigh the box, without opening it. 

Evidently, your probabilities $q(j)$ for the unopened box to weigh $j$ grams should be related to your probability assignment $p(i)$ for what you would find if you were to open it. In other words, your probability assignments $p(i)$ for a hypothetical measurement that is not performed, nevertheless constrain your beliefs about a different measurement $q(j)$ that is to be performed. 

More realistically, your probability assignment $q(j)$ would also depend on the properties of the scale used to weigh the box. Consider a box in which you are sure to find $i$ chocolates (that is, if you were to open it and count them) and let $r(j|i)$ be your probability that the scale would read `$j$' if such a box were placed upon it. The set of  $r(j|i)$ for all values of $i$ characterizes your beliefs about the functioning of the weighing scale. In general, then, your probabilities $q(j)$ for weighing any given box of chocolates should be constrained by a rule of the general form:
\eqn{ \label{eqn:generalrule}
q(j) = \mathcal{F}\left[ p(i), r(j|i) \right] \, .
}
This relation is not enforced by any Dutch book argument, rather, it is a normative addition to probability theory, enforced by World-coherence based on your previous dealings with boxes of chocolate and kitchen scales. 

In order to generalize this rule to hold for any measurement we might conceivably perform, whose outcomes we continue to label $j$, we must choose our hypothetical $i$-measurement to be much more thorough than merely counting the chocolates. It would have to be a measurement for which the probabilities $p(i)$ would suffice to constrain the probabilities of \tit{any} alternative measurement via some rule of the form \eqref{eqn:generalrule}; such a measurement is called \tit{informationally complete} (IC).

Thus, in our example, we might imagine for our $i$-measurement not just counting the chocolates, but simultaneously measuring their sizes, weighing them, taking their temperature, and so on. In general, depending on the choice of IC-measurement, the resulting form of the relation $\mathcal{F}$ could be rather complicated. This is especially true if the measurement somehow disturbs the system (for instance if chocolates get eaten during the process), for then the function $\mathcal{F}$ must correct for the effects of this disturbance. It is a remarkable feature of our classical theories that they always permit (in principle) an IC-measurement that is entirely non-disturbing.
Formally, `non-disturbing' means that, for the purpose of assigning the probabilities $q(j)$, the Agent is indifferent between the following two circumstances:\\
(i) the IC-measurement is performed, and its outcome $i$ is ignored;\\
(ii) the IC-measurement is not performed at all.\\
Therefore, the statement that the IC-measurement is `non-disturbing' amounts to the normative constraint 
\eqn{ \label{eqn:nondisturb}
q(j) = s(j) \, ,
}
where $s(j)$ is the probability for the result $j$ given that the IC-measurement is performed and the outcome $i$ ignored. Since an Agent is free to judge a measurement to be non-disturbing or not depending on past experience, \eqref{eqn:nondisturb} is an instance of World-coherence.

The form of $s(j)$ is constrained by a basic result of probability theory (derivable from Dutch book coherence) called the \tit{Law of Total Probability}:
\eqn{ \label{eqn:LTP}
s(j) = \zum{i}{} r(j|i)p(i) \, .
}
Combining this with \eqref{eqn:nondisturb}, we obtain the form of the normative rule \eqref{eqn:generalrule} applicable to non-disturbing IC-measurements:
\eqn{ \label{eqn:classurg}
q(j) = \zum{i}{}  r(j|i)p(i) \, .
} 
Although this looks superficially similar to the Law of Total Probability, note that whereas \eqref{eqn:LTP} is a consequence of Dutch book coherence, \eqref{eqn:classurg} depends also on \eqref{eqn:nondisturb} and so is contingent on World-coherence. In particular, an Agent will always adhere to \eqref{eqn:LTP}, but experience may give them reason to reject \eqref{eqn:classurg}. We will soon consider an alternative to \eqref{eqn:classurg} that applies when the Agent believes that the system in question is a quantum system.

First note that, since the rule \eqref{eqn:classurg} allows one to calculate $q(j)$ for \tit{any} measurement whatsoever on the system (by virtue of the $i$-measurement being informationally complete) it encodes the Agent's beliefs about all possible measurements on the system. By fixing the choice of IC-measurement according to some convention, which we call the \tit{reference measurement}, we can interpret the probabilities $p(i)$ that the Agent would assign to the reference measurement as representing the `state' of a system.

One of the features that distinguishes quantum mechanics from classical physics is that there is no IC-measurement for which the function $\mathcal{F}$ takes the simple form of \eqref{eqn:classurg}. There is, however, an informationally complete measurement that closely approximates it, called a \tit{symmetric informationally complete} POVM (SIC-POVM) \cite{DFS2020}. With this choice of reference measurement, the normative rule becomes:
\eqn{ \label{eqn:urgleichung}
q(j) = \zum{i}{d^2} \, \left[(d+1)p(i)-\frac{1}{d} \right]r(j|i) \, ,
}
which is called the \tit{Urgleichung}, and it can be derived straightforwardly from the Born rule \cite{QPLEX}. Here, $d$ is interpretable as the number of perfectly distinguishable states of the system (its `informational dimension') and also corresponds to the dimension of the system in the Hilbert space formulation of quantum theory. QBism proposes that \eqref{eqn:urgleichung} encodes the Agent's beliefs about all possible measurements on a system, when that system is judged to be quantum. Conversely, if experience leads the Agent to use a rule of the form \eqref{eqn:urgleichung} instead of \eqref{eqn:classurg}, we may take this as defining what we mean when we say the system `is quantum'. Using \eqref{eqn:urgleichung}, we can interpret the `quantum state' as meaning the set of probabilities $p(i)$ that an Agent would assign to the result $i$ of a hypothetical reference SIC-POVM on a quantum system.

The normative rule \eqref{eqn:urgleichung} is not to be understood as a replacement for the classical rule \eqref{eqn:classurg}. The latter remains useful in situations where there exists a measurement that is effectively non-disturbing and IC relative to the set of possible measurements the Agent can perform, which may occur when this set is sufficiently limited, say, by the precision of the Agent's apparatuses. However, when the possible measurements on a system become rich enough to exhibit quantum phenomena, the appropriate normative rule is \eqref{eqn:urgleichung}. That is why QBists say the latter is an \tit{addition} to the Agent's repertoire of normative rules, rather than a replacement of any existing rules. As an analogy, the theory of General Relativity does not replace Newtonian Gravity as a predictive tool, but rather supplements it in those domains where the latter no longer yields results that agree with experiments.

Although it is currently unknown whether SIC-POVMs exist for Hilbert spaces of all dimensions, QBists prefer to adopt SICs as the reference measurement of choice, because the simple form of \eqref{eqn:urgleichung} lends it to being taken as an axiom for a probabilistic reconstruction of quantum theory that does not assume the Hilbert space representation \cite{QPLEX}. Even if SICs do not always exist, one can still get a uniquely quantum (albeit more complicated) expression for \eqref{eqn:generalrule} by using other IC constructions. For instance, \tit{minimal} IC-POVMs, which have exactly $d^2$ outcomes, exist in all dimensions \cite{CFS2002,BFS2018}.

In the remaining two subsections, we will elaborate on two features that have remained relatively under-developed in QBist literature. The first is of a more technical nature and concerns what is meant by a `system', while the latter examines the QBist understanding of the Agent's perceptions and their body.

\subsubsection{Systems \label{Sec:systems}}

So far QBists have taken for granted that an Agent can identify and repeatedly address individual `systems' within their external World. Since there is no guarantee \tit{a priori} that different Agents will agree on what system is being addressed by a given measurement, we will capitalize `System' henceforth to indicate the System as judged by an individual Agent.

The problem of identifying the System can be framed as follows. Suppose an Agent is contemplating doing a measurement chosen from a set indexed by $k = 1,2,\dots,K$. Let $q_k(j)$ be the probabilities they assign to the results $j = 1,2,\dots,J_k$ of the $k$-th measurement. What can it mean to say that these $K$ measurements are all addressing the same System $S$ of dimension $d$? Usually QBists begin by \tit{stipulating} that this is the case, and then \tit{concluding} that there must exist a density matrix $\rho$ and a POVM $\{ E^{(k)}_j : j = 1,2,\dots,J_k\}$ for each $k$, such that the probability assignments may be written as:
\eqn{
q_k(j) = \tr{\rho E^{(k)}_j} \, .
}

Recalling our discussion of the previous section, we can equivalently express this condition as saying that there must exists an assignment of probabilities $p(i)$ to a hypothetical SIC-POVM with outcomes $i = 1,2,\dots,d^2$, and a set of $K$ conditional probability distributions $r_k(j|i)$, such that the agent's probability assignments may be written as:
\eqn{ \label{eqn:qbist_system}
q_k(j) &:=&  \zum{i}{d^2} \, \left[(d+1)p(i)-\frac{1}{d} \right]r_k(j|i) \qquad \forall \, k \, . \nonumber \\
&& 
}
Thus, the usual approach assumes that the Agent has some independent reasons for judging a set of measurements to be addressing a single System, and then imposes \eqref{eqn:qbist_system} as a necessary constraint on the Agent's probability assignments that follows from this belief. In the present work, we propose that this logic can also be inverted. For suppose that an Agent has independent reasons to assign the set of probability distributions $\{ q_k(j) : \forall k\}$, and they just so happen to notice that there exist valid distributions $p(i)$ and $\{ r_k(j|i) : \forall k\}$ such that the constraint \eqref{eqn:qbist_system} holds. Then the Agent is licensed to postulate that these measurements are all addressing the same System. 

This provides us with a means by which the ability to perform new measurements can lead to the discovery of new Systems that were not suspected prior to performing the measurements. This is important because, if Systems always had to be presupposed before we could talk about measurements upon them, then an Agent would have no means of discovering new Systems through the increased refinement of their measuring Apparatus. By allowing that the ability to perform new measurements may logically precede the identification of the Systems to which they are targeted, we enable QBism to account for the discovery of new Systems through the application of new Apparatuses. 

Finally, we remark on an interesting feature of the form of \eqref{eqn:qbist_system}. The probabilities $q_k(j)$ on the LHS are taken to refer to a single measurement Event. However, the RHS of the equation \eqref{eqn:qbist_system} `splits up' these probabilities into two parts. The first part, $p(i)$, is independent of the Agent's choice of measurement $k$, and so we may interpret it as assigning the `state' of the System independently of how it is measured. The second part, $r_k(j|i)$, encodes the Agent's possible experiences for any state of the System; as such we may interpret it as characterizing the response of an Apparatus independently of any particular state. Thus, the Event anticipated by the probabilities $q_k(j)$ is thereby conceptually dissected into the union of a System in a state $p(i)$ and an Apparatus delineated by $r_k(j|i)$, which we will shortly associate to the Agent's body.

\subsubsection{The body of the Agent}

The features of an Event contain information about the sensory body of the Agent. As an example, consider driving over a bump in the road. Here are two distinct but equally natural ways of describing the experience:\\ 

\noindent (1) `I was driving along on the road when I went over a bump'.\\

\noindent (2) `I was driving along in my car when the car went over a bump'.\\

In case (1) the driver's attention is on the road. Although driving a car, this Agent is not aware of the car as a separate external entity. Rather, impacts to the car, such as the bump, are perceived directly by the Agent; the car is, for all intents and purposes, an extension of the Agent's body. By contrast, the phrasing in example (2) makes it clear that the driver's attention is focused on the car itself, and the bump in the road is not directly felt, but is inferred through the jolting of the car; the latter is felt directly by the Agent's corporeal body seated in it. We have in effect two different Agents and hence two different personal Events, which may be stated more explicitly as:\\

\noindent Event 1: Agent 1 feels a bump in the road.\\

\noindent Event 2: Agent 2 feels a jolt in the car seat.\\

We chose this example because of its nearness to commonplace experience: most people have experienced driving over a bump in each of these two ways. Usually, as a driver, one is more aware of the road and less aware of the car, and so will tend to experience the car as an extension of their own sensory body, whereas a passenger will tend to be more aware of the movements of the car seat, and their sensory body effectively ends at their own skin. But much more exotic ways of experiencing the event could be imagined. A more sensitive and attentive driver might be able to directly feel not only the presence but also the precise shape of the bump; at the other extreme, a passenger deep in meditation might experience the bump not directly by the jolting of the seat but by the associated changes in their internal bodily state -- an increase in heart rate, slight movement of the organs, etc. Each way of experiencing the bump corresponds to a distinct placement of the boundary between the Agent and the World, part of which implicitly delineates the physical boundary of the Agent's sensory organs, which may be natural or artificial. In our present formalism, this choice enters in how we interpret the events $X^a$ to which we assign probabilities, that is, to what specific distinctive experiences the values $x \in X^a$ represent.  

For the purposes of giving a scientific but personalist description of any occurrence, we must make a definite choice of the Agent/World boundary (and hence of the Agent's effective body). Whether a choice is appropriate or inappropriate is to be judged primarily by the scientist whose experience is being described, for only they can know what they felt to be the boundaries of their own senses. If we wish to speak of Agents and Events in the abstract, without describing actual experiences and happenings, then we may be content so long as our choice of the Agent's sensory body is a plausible one (i.e.\ that the experiences thereby described are possible experiences for us in our role as scientists) and provided we remain consistent in this choice and do not move it around without explicitly acknowledging that we are doing so.

It is worth noting that our example of driving over a bump highlights the problem of objectivity in QBism. We have here something that would ordinarily be treated as a single unique event, namely, `the car went over the bump', and interpreted it instead as a variety of fundamentally distinct Agents and Events. Yet all of these fractions involve ostensibly the `same' physical entities -- a person, a car, and a bump. The question then arises as to how this conceptual unity comes about, that is, how it is possible to resolve the objective `person' and `event' out of the myriad possible perspectives of Agents and their Events. Whereas the usual approach in physics would presuppose an objective world and explain the different perspectives by `carving out' parts of it to serve as the perceiving subjects, QBism begins with a fundamental plurality of perspectives and seeks to derive some notion of objective concepts from them.

The present work does not aim at a general solution to the question of objectivity, but focuses on the problem of how to relate Agents to one another through the relative refinement of their Apparatuses. To the extent that the car in the example may be thought of as an Apparatus for sensing the bump in the road, our aim is to understand how the experiences of Agents like (1) and (2) in the example above must be formally related to one another. For although they are ostensibly different Agents experiencing different Events, it is evident that there is a continuity between them: Agent 1 is conceivable as resulting from an extension of Agent 2 via the means of the car. The latter is an external System probed by Agent 2 on the one hand, and on the other hand it is an Apparatus used by Agent 1. We aim to understand and formalize this kind of relationship for measurements on quantum systems. In particular, our goal is to understand in detail how the Agent's measurement Apparatus for making quantum measurements can be understood and formally described as a physical extension of the Agent.

\section{Extending the Agent \label{Sec:Extension}}

Suppose that the Agent is interested in doing measurements on a System $\mathcal{T}$, which we will refer to as the `target System', but that their repertoire of direct measurements on $\mathcal{T}$ is limited. Thus, they would like to find ways to extend their possibilities of interacting with $\mathcal{T}$ by involving another System $\mathcal{S}$, to serve as an Apparatus. In the following sections, we give an account of how the System $\mathcal{S}$ goes from being merely another System to becoming part of the perceptual Apparatus of the Agent, and as such, part of the Agent's `body'. 
We divide this process into two steps. The first step involves `tuning' System $\mathcal{S}$ to System $\mathcal{T}$ such that measurements on the former can be used to make inferences about hypothetical measurements on the latter. At this stage, however, $\mathcal{S}$ is still treated as a System external to the Agent, on which the Agent can perform measurements. The second step involves re-defining the boundary between Agent and World so as to include System $\mathcal{S}$ as part of the Agent, which we call the `extension of the Agent'. In the following subsections we describe each of these steps in detail.

\subsubsection{Tuning a prospective Apparatus}

Not any System can be used immediately as an Apparatus, for the simple reason that measurements on $\mathcal{S}$ do not necessarily have anything to do with measurements on $\mathcal{T}$. Intuitively, it must first be `tuned' -- there must exist some special composite action the Agent can take that we call `tuning $\mathcal{S}$ to $\mathcal{T}$', which encompasses any and all operations that need to be performed to ensure that the prospective Apparatus is in a position ready to measure the target System. 
For instance, if we aim to use a microscope to view microbes inside a drop of dew, we must first place the dew on a slide and position the slide on the microscope stage. Moreover, we need to understand how to operate the microscope, and perhaps to some extent how to model the microscope; all of this prior learning and activity is implicit in the action of tuning. This example already hints at the complexity involved in giving a full account of the process that makes a System suitable for use by an Agent as an Apparatus. This problem is discussed at length by Pickering \cite{Pickering_Mangle}, who describes the tuning process evocatively as a `dance of agency' that strives to attain a `capture of material agency':
\begin{quote}
As active, intentional beings, scientists tentatively construct some new machine. They then adopt a passive role, monitoring the performance of the machine to see whatever capture of material agency it might effect. Symmetrically, this period of human passivity is the period in which material agency actively manifests itself. Does the machine perform as intended? Has an intended capture of agency been effected? Typically the answer is no, in which case the response is another reversal of roles: human agency is once more active in a revision of modelling vectors, followed by another bout of human passivity and material performance, and so on. The dance of agency, seen asymmetrically from the human end, thus takes the form of a dialectic of resistance and accommodation, where resistance denotes the failure to achieve an intended capture of agency in practice, and accommodation an active human strategy of response to resistance, which can include revisions to goals and intentions as well as to the material form of the machine in question and to the human frame of gestures and social relations that surround it.
\end{quote}

Due to the evident difficulty in giving a characterization of the action of tuning that is both mathematically precise and conceptually general, we will sidestep the issue by focusing instead on the end product, that is, on characterizing the special relationship between $\mathcal{S}$ and $\mathcal{T}$ that is assumed to represent the outcome of a successful tuning process.

Intuitively, `tuning' is successful if it is possible to \tit{simulate} a direct measurement $Z$ on $\mathcal{T}$ by actually doing a direct measurement $Y$ on $\mathcal{S}$. We can formalize this idea in Hilbert space using the notion of a \tit{generalized dilation} of a POVM. First, let us represent the measurements $Z$ and $Y$ by (with a slight abuse of notation) the POVMs $\mathcal{Z} := \{Z_z : z \in Z \}$ and $\mathcal{Y} := \{ Y_y : y \in Y \}$, satisfying
\eqn{
\zum{z \in Z}{}\, Z_z &=& \mathds{1}^{\mathcal{T}} \nonumber \\
\zum{y \in Y}{}\, Y_y &=& \mathds{1}^{\mathcal{S}} \nonumber \\
}
with $\mathds{1}^{\mathcal{T}},\mathds{1}^{\mathcal{S}}$ the identity operators on the Hilbert spaces $\mathcal{H}^{\mathcal{S}},\mathcal{H}^{\mathcal{T}}$ of the respective Systems. We now define:\\

\tbf{Generalized dilation:} The POVM $\mathcal{Y}$ is called a generalized dilation of the POVM $\mathcal{Z}$ if both have the same number of possible results, and there exists an initial state $\sigma$ of $\mathcal{S}$ and a CPT map 
\eqn{
\Phi: \mathcal{L}(\mathcal{H}^{\mathcal{S}} \otimes \mathcal{H}^{\mathcal{T}}) \mapsto  \mathcal{L}(\mathcal{H}^{\mathcal{S}} \otimes \mathcal{H}^{\mathcal{T}}) \, ,
}
where $\mathcal{L}(\mathcal{H})$ denotes the space of linear operators on $\mathcal{H}$, such that
\eqn{ \label{eqn:gen_dilation}
\tr{\rho Z_z} = \tr{\Phi(\sigma \otimes \rho) (Y_y \otimes \mathds{1}^{\mathcal{T}})} \quad \forall y=z,
}
holds for all valid states $\rho$ of $\mathcal{T}$. If we define 
\eqn{
\rho' := \trc{\mathcal{T}}{\Phi(\sigma \otimes \rho)} \, ,
}
where $\trm{Tr}_{\mathcal{T}}$ is the partial trace over $\mathcal{H}^{\mathcal{T}}$, this condition simplifies to:
\eqn{ \label{eqn:gen_dilation_simple}
\tr{\rho Z_z} = \tr{\rho' Y_y} \qquad \forall y=z,
}
In other words, $\mathcal{Y}$ is a generalized dilation of $\mathcal{Z}$ if we can simulate $\mathcal{Z}$ by performing the following sequence of actions:\\
(i) preparing $\mathcal{S}$ in the state $\sigma$;\\
(ii) coupling $\mathcal{S}$ to $\mathcal{T}$ via the map $\Phi$; \\
(iii) performing $\mathcal{Y}$ on $\mathcal{S}$, and then discarding $\mathcal{S}$.\\ 
We will say that the measurement $Y$ is a generalized dilation of $Z$ whenever this is true of the POVMs representing these measurements. For future reference, we will refer to the sequence of actions (i)-(iii) as \tit{applying $\mathcal{S}$ to $\mathcal{T}$}.

Note that, in the special case where $\mathcal{Y}$ is a von-Neumann measurement, (i.e.\ its elements are orthogonal rank-1 projectors), our definition reduces to the standard notion of a `dilation' of a POVM. 

We can now define what characterizes the end goal of the `tuning' process:\\

\tbf{Tuned:} System $\mathcal{S}$ is \tit{tuned to} System $\mathcal{T}$ with respect to a set of measurements $\{ Y^b : b \in B \}$ on $\mathcal{S}$, if each member of this set can be paired with a measurement $Z^b$ on $\mathcal{T}$ such that $Y^b$ is a generalized dilation of $Z^b$.\\

It is worth pointing out that the above definitions can be formulated purely in terms of probabilities. This is important to the QBist, who seeks to express quantum theory purely in terms of constraints on an Agent's probability assignments. To see how this is achieved, note that the defining relation \eqref{eqn:gen_dilation_simple} of a generalized dilation can be neatly expressed as the constraint:
\eqn{ \label{eqn:stabilization}
P(z) = P(y) \qquad \forall y=z,
}
together with
\eqn{ \label{eqn:gendil2}
P(z) =  \zum{i}{d^2_{\mathcal{T}}} \, \left[(d_{\mathcal{T}}+1)p(i)-\frac{1}{d_{\mathcal{T}}} \right]r(z|i) \, ,
}
where $p(i),r(z|i)$ are the `SIC representations' of the state $\rho$ and POVM $\mathcal{Z}$, and $d_{\mathcal{T}}$ is the dimension of System $\mathcal{T}$; and 
\eqn{ \label{eqn:gendil3}
P(y) =  \zum{i}{d^2_{\mathcal{S}}} \, \left[(d_{\mathcal{S}}+1)p'(i)-\frac{1}{d_{\mathcal{S}}} \right]r'(y|i) \, ,
}
where $p'(i),r'(y|i)$ are the SIC representations of the state $\rho'$ and POVM $\mathcal{Y}$ and $d_{\mathcal{S}}$ the dimension of System $\mathcal{S}$. 

The conjunction of the constraints \eqref{eqn:stabilization},\eqref{eqn:gendil2},\eqref{eqn:gendil3}, when required to hold for all valid states $p(i)$ of $\mathcal{T}$, gives us a definition for the statement `$Y$ as a generalized dilation of $Z$' purely in terms of the Agent's probability assignments. We can therefore understand the statement `$\mathcal{S}$ is tuned to $\mathcal{T}$' as a statement about a constraint on the Agent's beliefs about the two Systems. Specifically, it means that the Agent believes that certain measurements on $\mathcal{S}$ are `just as good as', i.e.\ a proxy for, certain measurements on $\mathcal{T}$. We will therefore say that, under these conditions, the System $\mathcal{S}$ is a \tit{prospective Apparatus} for measuring $\mathcal{T}$. It is not yet a full Apparatus, because it is still regarded as an System external to the Agent.

Note that the condition \eqref{eqn:stabilization} can be thought of as a kind of non-contextuality assumption, namely, that the probability of result $y$ doesn't depend on whether it was measured as part of $Y$ or part of $Z$. Unlike standard measurement non-contextuality, however, the contexts being referred to here are not a pair of alternative measurements on the \tit{same} System, but rather a pair of measurements on two different Systems.

As an analogy, consider using an Ammeter to measure the current in a wire. The perceived position of the Ammeter needle is akin to the Event $Y^b$: it represents a direct measurement of the Ammeter, and only an indirect measurement of the current. It is a reliable indicator of the \tit{amount of current} you would measure if you were to try measuring it by grasping the electric wire with your hand (analogous to measuring $Z^b$). The measurement $Y^b$ is therefore a good \tit{proxy} for $Z^b$, in the restricted sense that you would assign the same probabilities to the different values of current `perceived' in each case (with the proviso that the Ammeter has resolution and operating range comparable to that of human hands). This is what is captured by the normative constraint \eqref{eqn:stabilization}. Nevertheless, the two are evidently quite different perceptual experiences and the correspondence only holds for one special feature of the respective Events, namely the quantified amount of current experienced. For other features, such as the level pain experienced, $Y^b$ is not a suitable proxy for $Z^b$. Incidentally, this example illustrates that the act of `tuning' a prospective Apparatus to a System is specific to some particular feature or features of interest, in this case the quantified current.

\subsubsection{Incorporating the Apparatus}

The action of tuning $\mathcal{S}$ to $\mathcal{T}$ is not itself sufficient to elevate $\mathcal{S}$ to an Apparatus regarded as part of the Agent. To attain this designation, we require more than just that measurements on $\mathcal{S}$ `simulate' or are a `proxy' for measurements on $\mathcal{T}$ -- we require that these become equivalent to \tit{direct} measurements of $\mathcal{T}$, as it were, `through' the System $\mathcal{S}$. To illustrate this idea, consider the commonplace act of trying on a new pair of spectacles. At first, one regards them at arm's length and is aware of the spectacles as an external System. The image that one perceives through their lenses is perceived as being `in the lenses', or emanating from the lenses of the spectacles. When one has worn the spectacles for a while, a transition occurs: one ceases to see the glass of the lenses, and ceases to see the images as being `in' the glass -- instead the images are given in perception directly, as it were, \tit{through} the glass. The spectacles, in this new mode of usage, are as good as being an extension of one's own eyes. We now formalize this transition, which we will call the `extension of the Agent'.

After tuning, the Agent still regards $\mathcal{S}$ as an external physical System on which measurements can be performed, and \tit{tuning} as an action that she can take on these Systems that places them into the relation expressed by \eqref{eqn:stabilization}. However, once the Agent is confident that \eqref{eqn:stabilization} holds, she may find it useful to abstract away from the details of the System $\mathcal{S}$ and its interactions with $\mathcal{T}$. This can be done, for instance, by automating the procedure and literally placing it inside a `black box'. When a dial on the box is set to `$b \in B$' and a button is pushed, the box automatically applies $\mathcal{S}$ to $\mathcal{T}$ so as to simulate $Z^b$ by performing $Y^b$, effortlessly and without requiring thought or further intervention by the Agent. At this point, the Agent trusts the functioning of the box enough to ignore its inner workings, namely, the System $\mathcal{S}$ and the detailed modeling of the coupling process. She can simply pick it up and start using it \tit{as if it were in fact a direct measurement $Z^b$ on the System $\mathcal{T}$}. We can formalize this shift in perception as follows:\\

\tbf{Extension of the Agent:} At any time, the Agent is free to postulate that the operation of using a tuned prospective Apparatus to measure $Y^b$ is \tit{literally equivalent} to making the direct measurement $Z^b$ on System $\mathcal{T}$. In doing so, she alters her set $\{X^a : a \in A \}$ of direct measurements to a new set that includes (but is perhaps not limited to) $\{ Z^b : b \in B \}$, where now the actions indexed by $b \in B$ are interpreted as `perform measurement $Z^b$ directly on System $\mathcal{T}$ using the Apparatus'.

Simultaneously, she ceases to regard the System $\mathcal{S}$ as a System; ceases to regard $\{ Y^b : b \in B\}$ as possible measurements; and ceases to regard tuning $\mathcal{S}$ to $\mathcal{T}$ as a meaningful action. These concepts are subsumed by, and implicit in, the set of new direct measurements $\{ Z^b : b \in B \}$. We say that the Agent has \tit{extended herself to incorporate the prospective Apparatus} $\mathcal{S}$ \tit{as the measurements} $\{ Z^b : b \in B \}$ on $\mathcal{T}$. Thereafter, the means by which the Agent performs the measurements $Z^b$ -- what was formerly known as System $\mathcal{S}$ -- is referred to simply as the Apparatus.\\

We note that in the definition above, it was not specified what is the set of measurements the Agent can actually perform after incorporating $\mathcal{S}$ as an Apparatus, except that it should include the $Z^b$. There are, in fact, two natural possibilities: either the $Z^b$ are an \tit{addition} to the Agent's pre-existing possible measurements on $\mathcal{T}$, which we denote $X^a$, or else they are a \tit{replacement} for them. The first case applies whenever the acquisition of the Apparatus amounts to the addition of a new tool to an existing toolset: merely having another tool does not in itself prevent the Agent from using the previously existing tools if they wish to. For instance, although in a certain sense a pair of spectacles `replaces' the wearer's eyes while in use, the spectacles can be worn or removed at will, according to convenience, and the possession of the spectacles does not preclude the option of using one's eyes without spectacles. This category extends to cases where the usage of the Apparatus is expensive or difficult. What is at stake here is rather what measurements are \tit{possible} to the Agent, however easy or difficult they may be to implement. The alternative case is where the physical changes necessary to employ the new Apparatus are such as to entail that the old measurements are effectively \tit{impossible} (except insofar as they can be achieved by the new Apparatus). Irreversible physical modifications are in this category. The distinction between these cases is captured in the following definitions (hereafter we employ the shorthand of e.g.\ $\{ X^a\}$ for the set $\{X^a : a \in A \}$):\\

\noindent \tbf{Inclusive extension:} The Agent's new set of possible measurements is $\{ X^a \} \cup \{ Z^b \}$;\\

\noindent \tbf{Exclusive extension:} The the Agent's new set of possible measurements is only $\{ Z^b \}$.\\

Of particular interest is a special case in which these two modes of extension are equivalent, which occurs when $\{ X^a \} = \emptyset $, that is, when the Agent has no \tit{a priori} means of measuring System $\mathcal{T}$, and can initially only access it indirectly through the prospective Apparatus $\mathcal{S}$. In other words, it may happen that the Agent notices that there exist a subset of measurements on $\mathcal{S}$ and preparations of $\mathcal{S}$ that can be interpreted as generalized dilations for some hypothetical preparations $p(i)$ of an entirely hypothetical System $\mathcal{T}$, in the sense that probability assignments can be found to satisfy the requisite constraints \eqref{eqn:stabilization},\eqref{eqn:gendil2},\eqref{eqn:gendil3}. The System  $\mathcal{T}$ is initially `hypothetical' in the sense that it is \tit{only} accessible via the newly discovered indirect measurements $\{ Z^b \}$ through  $\mathcal{S}$. So long as the latter remains an external System, the reality of $\mathcal{T}$ remains merely hypothetical. When $\mathcal{S}$ finally is incorporated as an Apparatus, the $\{ Z^b \}$ become the first \tit{direct} measurements of $\mathcal{T}$, and it becomes known to the Agent as a System proper. This shows how new Systems can be `discovered' in QBism through the refinement of Apparatuses, as was promised in Sec.\ \ref{Sec:systems}.

To see how the `extension of the Agent' results in the literal extension of the Agent's body, note that since the process of incorporating a prospective Apparatus involves losing some measurements and actions on $\mathcal{S}$ and gaining new ones on $\mathcal{T}$, it amounts to a shifting of the boundary between Agent and World. After its incorporation by the extension of the Agent, the System $\mathcal{S}$ no longer exists as an external entity, but is replaced by an Apparatus that performs `the same' measurements, only now directly rather than indirectly, and so is considered part of the Agent. The World has thus shrunk by losing a System, but the Agent has grown in gaining an Apparatus.

Finally, in referring to the extension as an act of `free postulation' by the Agent, we mean that there is no external criterion by which the Agent's decision to extend themselves can be judged as right or wrong; they cannot fall prey to a Dutch book argument by choosing to incorporate a prospective Apparatus, or not. This is because, as mentioned in Sec.\ \ref{Sec:Overview}, quantum theory does not prescribe how to demarcate the `Agent' from the `external World'; rather, a choice of such demarcation is a prerequisite for applying the tools of quantum theory that must come from considerations outside of that theory.

\section{Comparing Apparatuses \label{Sec:comparison}}

\begin{quote}
    ``Ultimately I view QBism as a quest to point to something
    in the world and say, that's intrinsic to the world. But I don't have a conclusive
answer yet. Let's take the point of view that quantum mechanics is a user's manual. A
user's manual for me. A philosopher will quickly say, well that's just instrumentalism.
`Instrumentalism' is always prefaced by a `just'. But that's jumping too quickly to
a conclusion. Because you can always ask -- you should always ask -- what is it about
the world that compels me to adopt this instrument rather than that instrument?" \\
-- Fuchs, \tit{On Participatory Realism} \cite{Fuchs_PR}
\end{quote} 

In this section, we will define what is meant by a `more accurate' or `more refined' measurement in QBism, in a way that makes no reference to supposed pre-existing `true values' of the measured quantity. We will see that the problem of evaluating the accuracy of an Apparatus can be framed as a simple decision problem, which can be formulated entirely in subjective terms. We will find that an Agent will strictly prefer to use one Apparatus over another, whenever the former can be used to emulate the latter by post-processing of its results, but not conversely. This defines a partial `preference ordering' among Apparatuses, which turns out to be equivalent to the the `degradability ordering' of statistical channels that has been extensively studied in the literature (see e.g.\ \cite{Rauh2017,Blackwell,Raginsky}); in particular it allows us to make a connection with the recently proposed resource theory of quantum measurements \cite{GUFF}. We are thereby able to define an Apparatus as more accurate or more refined whenever it is more resourceful. Finally, we use the partial ordering of Apparatuses given us by the resource theory to make a simple taxonomy of the different kinds of Agent extension that are possible, relative to the Agent's given initial abilities.

\subsubsection{What makes an Apparatus more accurate?}

To clarify this issue, we introduce the following decision problem. An Agent is asked to guess the result of a future Event $W$, to which they assign a prior probability $P(w)$. Moreover, before the Event $W$ occurs, the Agent is allowed to perform a single measurement on System $\mathcal{T}$, whose result may provide additional information about $W$. Specifically, we assume that the Agent assigns probabilities $\{ q_a(x) \}$ to the results of measurements $\{ X^a \}$ on $\mathcal{T}$, which are decomposed as
\eqn{
q_a(x) = \zum{w \in W}{} \, q_a(x|w)P(w) \,
}
where $\{ q_a(x|w) \}$ represents the probabilities that the Agent thinks they \tit{would} assign to $\{ X^a \}$ if they knew that $W=w$.
(Incidentally, probability assignments like the above, in which the Agent anticipates their own future probability assignments, play an important role in QBism. The Dutch book coherence of such assignments is a result known as the \tit{reflection principle} \cite{FS_reflection}).
Given the outcome $X^a=x$ of their chosen measurement, the Agent guesses the value of $W$ according to a strategy represented by a conditional probability function $v_a(w'|x)$ where $w'$ is the Agent's guess. Finally, the Event $W=w$ happens and the Agent receives a reward according to some utility function $u(w',w)$, which we assume to be real-valued but otherwise unconstrained.

One way to motivate this problem is from the point of view of biology: the Agent may be an organism that seeks to anticipate some feature of its environment, represented by $W$. To do so, the organism uses its senses (its Apparatus) to measure some proximal part of the environment (the System) in order to glean information that will help it anticipate $W$. The organism's guess $w'$ may be interpreted as its pre-emptive action in anticipation of the stimulus $W$, and the utility function describes how the organism is rewarded or punished according to how appropriate its action was in light of $w$. Another interpretation, more appropriate to the present work, is that $W$ quantifies some anticipated physical phenomenon, say the severity of a coming storm, and $w'$ represents a meteorologist's prediction, based on their readings of some Apparatus, say a barometer, that is applied to the local atmosphere (the System). The utility function then quantifies the scientist's reward for making an accurate prediction. 
These examples show that our decision problem is quite general in its applicability, but it also has limitations. In particular, we only consider the task of making predictions, disregarding any `costs' that might be associated with the Apparatus, such as the physiological cost to an organism of evolving a new appendage, or the monetary cost to a scientist of buying a more accurate instrument. In principle one could account for these by allowing the utility function to depend on more parameters; however, as we are here aiming for a level of generality that is not concerned with such particulars, we leave aside the issue of the costs of different Apparatuses to focus on comparing them strictly according to what measurements they make possible for the Agent in principle.

Returning to the decision problem in the abstract, our goal is to compute the Agent's maximum expected utility, optimized over all possible guessing strategies $v_a(w'|x)$, for each possible choice of measurement $X^a$. That is, we define the maximum expected utility given a measurement $X^a$ as:
\eqn{
&& u_{\trm{max}}(X^a) := \nonumber \\
&& \qquad \underset{v_a}{\trm{sup}} \, \zum{w,w' \in W}{}\, u(w',w)v_a(w'|x)q_a(x|w)P(w) \, . \nonumber \\
&&
}
In general, different measurements will perform better or worse depending on the choice of utility function, which may depend on the Agent's personal aims. However, sometimes one measurement has a higher maximum expected utility than another measurement for \tit{all} possible utility functions. This leads us to a natural definition of a partial ordering on the set of possible measurements:\\

\tbf{Preference ordering:} Given two possible measurements $Z^b$ and $X^{a}$ on $\mathcal{T}$, we say that $Z^b$ is \tit{preferred} to $X^{a}$, denoted $Z^b \geq X^{a}$, if $u_{\trm{max}}(Z^b) \geq u_{\trm{max}}(X^{a})$ for \tit{any} choice of utility function $u(w',w)$. We say it is \tit{strictly preferred} if the inequality is strict.\\

The preference ordering is enforced by the following principle: given a choice between two actions $a$ and $a'$, which are rewarded according to a utility function $u(a)$, the Agent should choose $a$ instead of $a'$ whenever $u(a) > u(a')$. This decision problem has been studied previously in the field of noisy channel comparisons (see e.g.\ \cite{Raginsky}). A key result of Blackwell \cite{Blackwell} establishes the following equivalence, rendered using our own notation:\\

\tbf{Theorem (Blackwell):} Given measurements on $\mathcal{T}$ represented by the random variables $Z^b$ and $X^{a}$, the following are equivalent:\\
(i) $Z^b$ is preferred to $X^{a}$, i.e.\ $Z^b \geq X^{a}$;\\
(ii) $q_{a}(x|w) = \zum{z}{} \, \lambda(x|z) q_b(z|w)$;\\

\noindent where $\lambda(x|z)$ is a conditional probability distribution that represents some stochastic mixing of the measurement $Z^{b}$. Thus, our `preference order' defined above is the same as the `Blackwell' or `degradability' order studied elsewhere\cite{Rauh2017,Blackwell,Raginsky}. In the present context, it has the following interpretation: an Agent prefers $Z^b$ to $X^{a}$ whenever it is possible to simulate $X^{a}$ by doing $Z^{b}$ and then applying the stochastic map defined by $\lambda(x|z)$ to the result $z$. The significant point is that, arguably, an Agent always has the ability to perform such a re-mapping of measurement results, which in the quantum information literature is referred to as a \tit{classical post-processing}. 

Why is this so? It is natural to assume that, on receiving the result $Z^b=z$, the Agent can invent several new fictional results $z_1,z_2,\dots, z_M$, and manufacture a number $m \in \{1,2,\dots,M \}$ drawn from an arbitrary distribution $\lambda(m)$, and then declare the result of the measurement to be $z_m$ instead of $z$. Furthermore, given any set of distinct results $z,z',z'',\dots \in Z^b$, it is natural to assume that the Agent can conflate these by interpreting all results $z,z',z'',\dots$ as a single result $z$. It can be shown (a proof is in Ref.\ \cite{GUFF}) that the ability to perform these `free' transformations of the result $z$ is sufficient to enable to Agent to perform an arbitrary stochastic map $\lambda(x|z)$ on $z$. We therefore postulate that the sets of measurements the Agent can actually perform must be closed under arbitrary classical post-processing of their results. 

This postulate allows us to make an immediate connection to the ``\tit{Resource Theory of Quantum Measurements}" described by Guff et al. \cite{GUFF}. The following definitions and statements are implied by the results of that work:\\
1. The \tit{equivalence class} of any single measurement $X^a$ is defined as those measurements related to it by a reversible classical post-processing. Similarly, the equivalence class of a set of measurements $\{ X^a \}$ is the union of the equivalence classes of its individual members. \\
2. If $Z^b \geq X^{a}$ for two non-equivalent measurements, then the POVM that represents $X^{a}$ can be obtained by a classical post-processing of the results of the POVM that represents $Z^{b}$. It follows that $Z^{b}$ is \tit{more resourceful} than $X^{a}$ in the formal sense of Ref.\ \cite{GUFF}. Similarly, a set $\{ Z^{b} \}$ is more resourceful than another set $\{ X^a \}$ if every member of the latter can be obtained by a classical post-processing of a member of $\{ Z^{b} \}$.\\
3. The \tit{minimally informative measurement} $\mathcal{I}$ is defined as a single-result measurement for which the Agent always assigns unit probability. (Dutch book coherence guarantees that there always is such a proposition to which the Agent necessarily ascribes probability 1; it may be loosely interpreted as the proposition that `out of all things considered possible, one will occur'.) The equivalence class of $\mathcal{I}$ is the unique \tit{minimally resourceful} equivalence class: it is less resourceful than every other equivalence class of measurements. An Agent who can only perform measurements equivalent to $\mathcal{I}$ has nothing to lose by extending themselves to incorporate a new Apparatus.\\
4. A \tit{maximally resourceful} equivalence class is one for which there is no more resourceful equivalence class of measurements. 
Therefore a maximally resourceful equivalence class cannot be obtained by classical post-processing of any measurements from another equivalence class. In classical theory (taken to be the limiting case in which all measurements are represented by POVMs that are diagonal in a fixed basis) there is a unique maximally resourceful equivalence class, and hence a unique `best possible Apparatus'. This would be, for instance, the hypothetical Apparatus presumably used by Laplace's Demon to measure the positions and momenta of all fundamental classical particles to arbitrary precision. In quantum theory, there exists a distinct maximally resourceful equivalence class for every distinct POVM whose elements are all rank-1 (not necessarily orthogonal), because such POVMs cannot be transformed into one another by classical post-processing. Quantum theory therefore admits a rich abundance of distinct maximally resourceful Apparatuses. These include the equivalence classes of the various Von-Neumann measurements, as well as the equivalence classes of the various SIC-POVMs.\\

\subsubsection{A taxonomy of extensions}

Consider an Agent who initially is only able to perform a set of measurements $\{ X^a \}$ directly on System $\mathcal{T}$, and can do another set of measurements $\{ Y^b \}$ directly on a prospective Apparatus $\mathcal{S}$. After performing a extension of the Agent as described in Sec.\ \ref{Sec:Extension}, the latter are transformed into a new set $\{ Z^b  \}$ of direct measurements on $\mathcal{T}$. We argued in Sec.\ \ref{Sec:Extension} that the extension refers to a physical addition to the Agent's body in the sense that a System $\mathcal{S}$ previously belonging to the external World is subsumed as part of the Agent. However, commonplace examples show that a physical extension does not always correspond to an extension of the Agent's sensory powers: a badly designed pair of spectacles can just as well inhibit the Agent's vision as assist it. This is formally the case whenever the newly acquired set of POVMs is strictly less resourceful than the Agent's originally possible measurements, i.e.\ the case when $\{Z^b \} <  \{X^a \}$. More generally, we can identify four distinct cases of interest:
\begin{itemize}
	\item \tbf{Downgrade:} $\{Z^b\} < \{ X^a\}$;
	\item \tbf{Upgrade:} $\{Z^b\} > \{ X^a\}$;
	\item \tbf{Duplicate:} $\{Z^b\} = \{ X^a\}$;
	\item \tbf{Innovation}: None of the above \\ (denoted $\{Z^b\} \neq \{ X^a\}$)
\end{itemize}

When we take into account whether the mode of incorporation is Inclusive or Exclusive according to the definitions given in Sec.\ \ref{Sec:Extension}, we obtain 8 distinct classes, shown in Table \ref{table}. In the Table, `Final measurements' indicates the equivalence class of measurements after incorporating the Apparatus, and the final column shows how it compares to the Agent's original measurements $\{ X^a \}$. We use the notation `$\neq$' to indicate when two equivalence classes are not comparable by the ordering relation, that is, where each set contains measurements that cannot be obtained from the other set by classical post-processing.

\begin{table*}[ht] 
\caption{Types of extension by resourcefulness \label{table}}
\centering
\begin{tabular}{p{0.1\linewidth}|p{0.15\linewidth}|p{0.15\linewidth}|p{0.15\linewidth}|}
 \cline{2-4}
 & \multicolumn{1}{ c| }{Case} & \multicolumn{1}{ c| }{ Final measurements } & \multicolumn{1}{ c| }{ Comparison to $\{ X^a \}$} \\ 
 \cline{1-4}
\multicolumn{1}{ |c| }{Downgrade, } &  \multicolumn{1}{ c| }{Exclusive } & \multicolumn{1}{ c| }{ $\{ Z^b \}$} & \multicolumn{1}{ c| }{ $<$ }   \\ 
\cline{2-4}
\multicolumn{1}{ |c| }{$\{ Z^b \} < \{ X^a \}$ } &  \multicolumn{1}{ c| }{Inclusive } & \multicolumn{1}{ c| }{ $\{ X^a \}$} & \multicolumn{1}{ c| }{ $=$ }   \\ 
\cline{1-4}
\multicolumn{1}{ |c| }{Duplicate, } &  \multicolumn{1}{ c| }{Exclusive } & \multicolumn{1}{ c| }{ $\{ X^a \}$} & \multicolumn{1}{ c| }{ $=$ }   \\ 
\cline{2-4}
\multicolumn{1}{ |c| }{$\{ Z^b \} = \{ X^a \}$ } &  \multicolumn{1}{ c| }{Inclusive } & \multicolumn{1}{ c| }{ $\{ X^a \}$} & \multicolumn{1}{ c| }{ $=$ }   \\ 
\cline{1-4}
\multicolumn{1}{ |c| }{Upgrade, } &  \multicolumn{1}{ c| }{Exclusive } & \multicolumn{1}{ c| }{ $\{ Z^b \}$} & \multicolumn{1}{ c| }{ $>$ }   \\ 
\cline{2-4}
\multicolumn{1}{ |c| }{$\{ Z^b \} > \{ X^a \}$ } &  \multicolumn{1}{ c| }{Inclusive } & \multicolumn{1}{ c| }{ $\{ Z^b \}$} & \multicolumn{1}{ c| }{ $>$ }   \\ 
\cline{1-4}
\multicolumn{1}{ |c| }{Innovation, } &  \multicolumn{1}{ c| }{Exclusive } & \multicolumn{1}{ c| }{ $\{ Z^b \}$} & \multicolumn{1}{ c| }{ $\neq$ }   \\ 
\cline{2-4}
\multicolumn{1}{ |c| }{$\{ Z^b \} \neq \{ X^a \}$ } &  \multicolumn{1}{ c| }{Inclusive } & \multicolumn{1}{ c| }{ $\{ X^a, \, Z^b \}$} & \multicolumn{1}{ c| }{ $\neq$ }   \\ 
\cline{1-4}
\end{tabular}
\end{table*}

It may be immediately observed that Duplicate extensions and the Inclusive Downgrade extension are all equivalent. This is intuitive as can be seen by the following example. If you have a barometer, then you are indifferent to any of the following: acquiring an additional identical barometer; replacing your barometer with an identical one; and acquiring an additional less accurate barometer. It is little surprise that an Exclusive Downgrade is the only strictly non-preferred extension. By contrast, for an Upgrade it does not matter whether it is Inclusive or Exclusive. 

Perhaps the most interesting cases are the Innovations. Although not explicitly stated in the Table, an Inclusive Innovation $\geq$ Exclusive Innovation, as can be seen by comparing their sets of Final measurements. It can also be deduced that an Inclusive Innovation is never less resourceful than $\{ X^a\}$, and so is always preferred to an Exclusive Downgrade. Beyond this, little can be said about them on the basis of the resource-theory alone; other considerations must be taken into account, such as the Agent's particular goals. For instance, changing Apparatuses by an Exclusive Innovation requires the Agent to forfeit at least some measurements that were previously possible in order to obtain others that were previously impossible. If the Agent has no use for the new measurements and desperately needed the ones that were forfeit, the change is just as detrimental as an Exclusive Downgrade. On the other hand, if the newly gained measurements are useful and the forfeited measurements were not important to the Agent, then the change is as good as an Upgrade.

\section{Conclusions and outlook \label{Sec:conclusions}}

We began this work by pointing out the ambiguous character of scientific instruments, asking: how can they be regarded both as external Systems to be tuned and calibrated, and yet also be regarded as an extension of the scientist's senses? This would pose no great difficulty if we could interpret ``extension of the scientist's senses" as merely a figure of speech, and insist that the instruments are \tit{always} external to their users, as traditionally-minded physicists might insist. But this avenue is not available to QBism, one of whose tenets is that the instrument is literally a part of the Agent using it.

This brings out most clearly that the notion of `Agent' in QBism is not confined to an individual person, which is a point that QBists have emphasized before \cite{FAQBism}. What has remained mysterious is precisely how QBism delineates the boundary between Agent and World in their formalism, and how this boundary can move. The present work has achieved more clarity on these points.

To summarize, we have seen that the boundary between Agent and World is formally delineated by the set of possible direct measurements that an Agent can perform, and the Systems to which they are targeted (which depends on the Agent's probability assignments to the measurements). To specify these details is also to specify the Agent's sensory body, at least in its quantifiable aspects that are relevant for doing scientific measurements. Furthermore we have seen how this boundary can move via the two-step process of \tit{tuning} and \tit{extension of the Agent}. Tuning is an interactive process between Agent and an external System that establishes necessary conditions for the System to be regarded as an Apparatus. Once these are met, the Agent may extend themselves by mapping details about the prospective Apparatus into the Agent's set of possible direct measurements, thereby incorporating it.

Although there is no fundamental rule that governs the placement of the boundary \tit{a priori}, there are rules about how it can move once it has been placed. This in turn allows us to establish a continuity between an Agent before and after incorporating an Apparatus: the two can be related by showing that the possible measurements of the latter can be obtained from the former via the two-step process we have described.

As an addendum, we note that the process of extension can also be carried out in reverse. That is, at any moment, an Agent might choose to `deconstruct' a measurement $Z^b$ on System $\mathcal{T}$ by postulating the existence of an auxiliary System $\mathcal{D}$ such that measurement $W^b$ on $\mathcal{D}$ is a proxy for $Z^b$. The Agent may then choose to cease to regard the latter as a direct measurement on $\mathcal{T}$, and instead interpret it as being \tit{in fact} the result of measuring $W^b$ on $\mathcal{D}$. Running the process in this direction is not unique, as there are infinite choices of $\mathcal{D}$ and $W^b$ that would do the trick. Nevertheless, as before, the Agent is free to postulate one of these as being the one that correctly describes the situation, and this entails a redefinition of the boundary between Agent and World. This may be useful in describing what happens when an Agent's Apparatus breaks and ceases to function as such, thereby becoming conspicuous to the Agent as an external System.

By defining the rules of Agent extension, we have been able to clarify several points. First, we were able to establish a connection to the resource theory of quantum measurements, which allowed us to give a formal characterization of why an Agent would consider one Apparatus superior to another for the purposes of measurement, without appealing to pre-existing `true values' of quantities. Secondly, we have doubled down on the QBist notion that an Agent using an Apparatus `directly experiences' the results given through the Apparatus. Thus, QBism accommodates the idea that a sufficiently practiced scientist using an electron microscope to measure atoms might be said to literally `sense atoms' and not merely be making inferences about them as abstract or hypothetical entities \footnote{The word `sense' may need qualification here, since the nature of the experience still bears the indelible mark of the equipment through which the atoms are `sensed'. For this kind of sensing-through-equipment we might borrow the word \tit{Umsicht} from Heidegger.}. 

Finally, a caveat: although we have stressed that the placement of the Agent-World boundary is not subject to any fundamental constraints, this is not to say that its application to a given actual experiment or example is always appropriate. For one thing, all prospective parts of an Agent should be tuned to one another, and as discussed in Sec.\ \ref{Sec:Extension} we have left out the details of what this tuning process involves. Evidently the physical aspects of an Agent's interactions with an external System will play a role in determining how easy or difficult it is for the Agent to tune that System; there is a reason why it is easy to pick up a cane and start using it as an extension of one's arm, while using a chair or a piece of string for the same purpose can be difficult or impossible. To account for such differences, an in-depth analysis of the tuning process would be required.

\acknowledgments
I thank R. Schack, J. DeBrota, B. Stacey and C.A. Fuchs for valuable discussions. This work is based on ideas first presented at the QIRIF conference in V\"{a}xj\"{o}, Sweden (2019), and further developed based on discussions with the attendees of the Phenomenological Approaches to Physics conference at Stony Brook University, NY (2019). I am indebted to the organizers of both conferences. This work was supported in part by the John E. Fetzer Memorial Trust.


\begin{thebibliography}{23}%
\makeatletter
\providecommand \@ifxundefined [1]{%
 \@ifx{#1\undefined}
}%
\providecommand \@ifnum [1]{%
 \ifnum #1\expandafter \@firstoftwo
 \else \expandafter \@secondoftwo
 \fi
}%
\providecommand \@ifx [1]{%
 \ifx #1\expandafter \@firstoftwo
 \else \expandafter \@secondoftwo
 \fi
}%
\providecommand \natexlab [1]{#1}%
\providecommand \enquote  [1]{``#1''}%
\providecommand \bibnamefont  [1]{#1}%
\providecommand \bibfnamefont [1]{#1}%
\providecommand \citenamefont [1]{#1}%
\providecommand \href@noop [0]{\@secondoftwo}%
\providecommand \href [0]{\begingroup \@sanitize@url \@href}%
\providecommand \@href[1]{\@@startlink{#1}\@@href}%
\providecommand \@@href[1]{\endgroup#1\@@endlink}%
\providecommand \@sanitize@url [0]{\catcode `\\12\catcode `\$12\catcode
  `\&12\catcode `\#12\catcode `\^12\catcode `\_12\catcode `\%12\relax}%
\providecommand \@@startlink[1]{}%
\providecommand \@@endlink[0]{}%
\providecommand \url  [0]{\begingroup\@sanitize@url \@url }%
\providecommand \@url [1]{\endgroup\@href {#1}{\urlprefix }}%
\providecommand \urlprefix  [0]{URL }%
\providecommand \Eprint [0]{\href }%
\providecommand \doibase [0]{http://dx.doi.org/}%
\providecommand \selectlanguage [0]{\@gobble}%
\providecommand \bibinfo  [0]{\@secondoftwo}%
\providecommand \bibfield  [0]{\@secondoftwo}%
\providecommand \translation [1]{[#1]}%
\providecommand \BibitemOpen [0]{}%
\providecommand \bibitemStop [0]{}%
\providecommand \bibitemNoStop [0]{.\EOS\space}%
\providecommand \EOS [0]{\spacefactor3000\relax}%
\providecommand \BibitemShut  [1]{\csname bibitem#1\endcsname}%
\let\auto@bib@innerbib\@empty
\bibitem [{\citenamefont {Fuchs}\ \emph {et~al.}(2014)\citenamefont {Fuchs},
  \citenamefont {Mermin},\ and\ \citenamefont {Schack}}]{FDR2014}%
  \BibitemOpen
  \bibfield  {author} {\bibinfo {author} {\bibfnamefont {C.~A.}\ \bibnamefont
  {Fuchs}}, \bibinfo {author} {\bibfnamefont {N.~D.}\ \bibnamefont {Mermin}}, \
  and\ \bibinfo {author} {\bibfnamefont {R.}~\bibnamefont {Schack}},\ }\href
  {\doibase 10.1119/1.4874855} {\bibfield  {journal} {\bibinfo  {journal}
  {American Journal of Physics}\ }\textbf {\bibinfo {volume} {82}},\ \bibinfo
  {pages} {749} (\bibinfo {year} {2014})},\ \Eprint
  {http://arxiv.org/abs/https://doi.org/10.1119/1.4874855}
  {https://doi.org/10.1119/1.4874855} \BibitemShut {NoStop}%
\bibitem [{\citenamefont {Fuchs}(2017{\natexlab{a}})}]{Fuchs2017}%
  \BibitemOpen
  \bibfield  {author} {\bibinfo {author} {\bibfnamefont {C.~A.}\ \bibnamefont
  {Fuchs}},\ }\href {https://arxiv.org/abs/1705.03483} {\bibfield  {journal}
  {\bibinfo  {journal} {Mind and Matter}\ }\textbf {\bibinfo {volume} {15}},\
  \bibinfo {pages} {245} (\bibinfo {year} {2017}{\natexlab{a}})},\ \Eprint
  {http://arxiv.org/abs/arXiv:1705.03483} {arXiv:1705.03483} \BibitemShut
  {NoStop}%
\bibitem [{\citenamefont {Wheeler}(1982)}]{Wheeler}%
  \BibitemOpen
  \bibfield  {author} {\bibinfo {author} {\bibfnamefont {J.}~\bibnamefont
  {Wheeler}},\ }\href@noop {} {\emph {\bibinfo {title} {Mind in Nature: Nobel
  Conference XVII, Gustavus Adolphus College, St. Peter, Minnesota}}}\
  (\bibinfo  {publisher} {Harper \& Row, San Francisco},\ \bibinfo {year}
  {1982})\ pp.\ \bibinfo {pages} {1--23}\BibitemShut {NoStop}%
\bibitem [{\citenamefont {Fuchs}(2017{\natexlab{b}})}]{FuchsNWB}%
  \BibitemOpen
  \bibfield  {author} {\bibinfo {author} {\bibfnamefont {C.~A.}\ \bibnamefont
  {Fuchs}},\ }\href@noop {} {\bibfield  {journal} {\bibinfo  {journal} {Mind
  and Matter}\ }\textbf {\bibinfo {volume} {15}},\ \bibinfo {pages} {245}
  (\bibinfo {year} {2017}{\natexlab{b}})}\BibitemShut {NoStop}%
\bibitem [{\citenamefont {Pearl}(2009)}]{Pearl}%
  \BibitemOpen
  \bibfield  {author} {\bibinfo {author} {\bibfnamefont {J.}~\bibnamefont
  {Pearl}},\ }\href@noop {} {\emph {\bibinfo {title} {Causality}}}\ (\bibinfo
  {publisher} {Cambridge University Press},\ \bibinfo {year}
  {2009})\BibitemShut {NoStop}%
\bibitem [{\citenamefont {Pienaar}(2020)}]{PIEQBcausal}%
  \BibitemOpen
  \bibfield  {author} {\bibinfo {author} {\bibfnamefont {J.}~\bibnamefont
  {Pienaar}},\ }\href {\doibase 10.1103/PhysRevA.101.012104} {\bibfield
  {journal} {\bibinfo  {journal} {Phys. Rev. A}\ }\textbf {\bibinfo {volume}
  {101}},\ \bibinfo {pages} {012104} (\bibinfo {year} {2020})}\BibitemShut
  {NoStop}%
\bibitem [{\citenamefont {James}(1911)}]{WJ}%
  \BibitemOpen
  \bibfield  {author} {\bibinfo {author} {\bibfnamefont {W.}~\bibnamefont
  {James}},\ }\href@noop {} {\emph {\bibinfo {title} {Some Problems of
  Philosophy}}}\ (\bibinfo  {publisher} {Cambridge, MA and London: Harvard
  University Press, 1979.},\ \bibinfo {year} {1911})\BibitemShut {NoStop}%
\bibitem [{\citenamefont {Whitehead}(1929)}]{Whitehead}%
  \BibitemOpen
  \bibfield  {author} {\bibinfo {author} {\bibfnamefont {A.~N.}\ \bibnamefont
  {Whitehead}},\ }\href@noop {} {\emph {\bibinfo {title} {Process and
  Reality}}}\ (\bibinfo  {publisher} {New York: Macmillan},\ \bibinfo {year}
  {1929})\BibitemShut {NoStop}%
\bibitem [{\citenamefont {Merleu-Ponty}(1945)}]{MP}%
  \BibitemOpen
  \bibfield  {author} {\bibinfo {author} {\bibfnamefont {M.}~\bibnamefont
  {Merleu-Ponty}},\ }\href@noop {} {\emph {\bibinfo {title} {Phénoménologie
  de la perception / Phenomenology of Perception}}}\ (\bibinfo  {publisher}
  {Paris: Gallimard},\ \bibinfo {year} {1945})\ \bibinfo {note} {(trans. Donald
  Landes, London: Routledge, 2012.)}\BibitemShut {NoStop}%
\bibitem [{\citenamefont {Jeffrey}(2004)}]{Jeffreys}%
  \BibitemOpen
  \bibfield  {author} {\bibinfo {author} {\bibfnamefont {R.}~\bibnamefont
  {Jeffrey}},\ }\href@noop {} {\emph {\bibinfo {title} {Subjective Probability:
  The Real Thing}}},\ Subjective Probability: The Real Thing\ (\bibinfo
  {publisher} {Cambridge University Press},\ \bibinfo {year}
  {2004})\BibitemShut {NoStop}%
\bibitem [{\citenamefont {Fuchs}\ and\ \citenamefont
  {Schack}(2012)}]{FS_reflection}%
  \BibitemOpen
  \bibfield  {author} {\bibinfo {author} {\bibfnamefont {C.~A.}\ \bibnamefont
  {Fuchs}}\ and\ \bibinfo {author} {\bibfnamefont {R.}~\bibnamefont {Schack}},\
  }\enquote {\bibinfo {title} {Bayesian conditioning, the reflection principle,
  and quantum decoherence},}\ in\ \href {\doibase 10.1007/978-3-642-21329-8_15}
  {\emph {\bibinfo {booktitle} {Probability in Physics}}},\ \bibinfo {editor}
  {edited by\ \bibinfo {editor} {\bibfnamefont {Y.}~\bibnamefont
  {Ben-Menahem}}\ and\ \bibinfo {editor} {\bibfnamefont {M.}~\bibnamefont
  {Hemmo}}}\ (\bibinfo  {publisher} {Springer Berlin Heidelberg},\ \bibinfo
  {address} {Berlin, Heidelberg},\ \bibinfo {year} {2012})\ pp.\ \bibinfo
  {pages} {233--247}\BibitemShut {NoStop}%
\bibitem [{\citenamefont {DeBrota}\ \emph {et~al.}(2020)\citenamefont
  {DeBrota}, \citenamefont {Fuchs},\ and\ \citenamefont {Stacey}}]{DFS2020}%
  \BibitemOpen
  \bibfield  {author} {\bibinfo {author} {\bibfnamefont {J.~B.}\ \bibnamefont
  {DeBrota}}, \bibinfo {author} {\bibfnamefont {C.~A.}\ \bibnamefont {Fuchs}},
  \ and\ \bibinfo {author} {\bibfnamefont {B.~C.}\ \bibnamefont {Stacey}},\
  }\href {\doibase 10.1103/PhysRevResearch.2.013074} {\bibfield  {journal}
  {\bibinfo  {journal} {Phys. Rev. Research}\ }\textbf {\bibinfo {volume}
  {2}},\ \bibinfo {pages} {013074} (\bibinfo {year} {2020})}\BibitemShut
  {NoStop}%
\bibitem [{\citenamefont {Appleby}\ \emph {et~al.}(2017)\citenamefont
  {Appleby}, \citenamefont {Fuchs}, \citenamefont {Stacey},\ and\ \citenamefont
  {Zhu}}]{QPLEX}%
  \BibitemOpen
  \bibfield  {author} {\bibinfo {author} {\bibfnamefont {M.}~\bibnamefont
  {Appleby}}, \bibinfo {author} {\bibfnamefont {C.~A.}\ \bibnamefont {Fuchs}},
  \bibinfo {author} {\bibfnamefont {B.~C.}\ \bibnamefont {Stacey}}, \ and\
  \bibinfo {author} {\bibfnamefont {H.}~\bibnamefont {Zhu}},\ }\href {\doibase
  10.1140/epjd/e2017-80024-y} {\bibfield  {journal} {\bibinfo  {journal} {The
  European Physical Journal D}\ }\textbf {\bibinfo {volume} {71}},\ \bibinfo
  {pages} {197} (\bibinfo {year} {2017})}\BibitemShut {NoStop}%
\bibitem [{\citenamefont {Caves}\ \emph {et~al.}(2002)\citenamefont {Caves},
  \citenamefont {Fuchs},\ and\ \citenamefont {Schack}}]{CFS2002}%
  \BibitemOpen
  \bibfield  {author} {\bibinfo {author} {\bibfnamefont {C.~M.}\ \bibnamefont
  {Caves}}, \bibinfo {author} {\bibfnamefont {C.~A.}\ \bibnamefont {Fuchs}}, \
  and\ \bibinfo {author} {\bibfnamefont {R.}~\bibnamefont {Schack}},\ }\href
  {\doibase 10.1063/1.1494475} {\bibfield  {journal} {\bibinfo  {journal}
  {Journal of Mathematical Physics}\ }\textbf {\bibinfo {volume} {43}},\
  \bibinfo {pages} {4537} (\bibinfo {year} {2002})},\ \Eprint
  {http://arxiv.org/abs/https://doi.org/10.1063/1.1494475}
  {https://doi.org/10.1063/1.1494475} \BibitemShut {NoStop}%
\bibitem [{\citenamefont {DeBrota}\ \emph {et~al.}(2018)\citenamefont
  {DeBrota}, \citenamefont {Fuchs},\ and\ \citenamefont {Stacey}}]{BFS2018}%
  \BibitemOpen
  \bibfield  {author} {\bibinfo {author} {\bibfnamefont {J.~B.}\ \bibnamefont
  {DeBrota}}, \bibinfo {author} {\bibfnamefont {C.~A.}\ \bibnamefont {Fuchs}},
  \ and\ \bibinfo {author} {\bibfnamefont {B.~C.}\ \bibnamefont {Stacey}},\
  }\href {https://arxiv.org/abs/1812.08762} {\  (\bibinfo {year} {2018})},\
  \bibinfo {note} {eprint arXiv:1812.08762}\BibitemShut {NoStop}%
\bibitem [{\citenamefont {Pickering}(1996)}]{Pickering_Mangle}%
  \BibitemOpen
  \bibfield  {author} {\bibinfo {author} {\bibfnamefont {A.}~\bibnamefont
  {Pickering}},\ }\href {\doibase 10.2307/3106908} {\bibfield  {journal}
  {\bibinfo  {journal} {Bibliovault OAI Repository, the University of Chicago
  Press}\ }\textbf {\bibinfo {volume} {38}} (\bibinfo {year} {1996}),\
  10.2307/3106908}\BibitemShut {NoStop}%
\bibitem [{\citenamefont {Fuchs}(2016)}]{Fuchs_PR}%
  \BibitemOpen
  \bibfield  {author} {\bibinfo {author} {\bibfnamefont {C.~A.}\ \bibnamefont
  {Fuchs}},\ }\href {https://arxiv.org/abs/1601.04360} {\  (\bibinfo {year}
  {2016})},\ \Eprint {http://arxiv.org/abs/arXiv:1601.04360} {arXiv:1601.04360}
  \BibitemShut {NoStop}%
\bibitem [{\citenamefont {Rauh}\ \emph {et~al.}(2017)\citenamefont {Rauh},
  \citenamefont {Banerjee}, \citenamefont {Olbrich}, \citenamefont {Jost},
  \citenamefont {Bertschinger},\ and\ \citenamefont {Wolpert}}]{Rauh2017}%
  \BibitemOpen
  \bibfield  {author} {\bibinfo {author} {\bibfnamefont {J.}~\bibnamefont
  {Rauh}}, \bibinfo {author} {\bibfnamefont {P.~K.}\ \bibnamefont {Banerjee}},
  \bibinfo {author} {\bibfnamefont {E.}~\bibnamefont {Olbrich}}, \bibinfo
  {author} {\bibfnamefont {J.}~\bibnamefont {Jost}}, \bibinfo {author}
  {\bibfnamefont {N.}~\bibnamefont {Bertschinger}}, \ and\ \bibinfo {author}
  {\bibfnamefont {D.}~\bibnamefont {Wolpert}},\ }\href {\doibase
  10.3390/e19100527} {\bibfield  {journal} {\bibinfo  {journal} {Entropy}\
  }\textbf {\bibinfo {volume} {19}} (\bibinfo {year} {2017}),\
  10.3390/e19100527}\BibitemShut {NoStop}%
\bibitem [{\citenamefont {Blackwell}(1953)}]{Blackwell}%
  \BibitemOpen
  \bibfield  {author} {\bibinfo {author} {\bibfnamefont {D.}~\bibnamefont
  {Blackwell}},\ }\href {\doibase 10.1214/aoms/1177729032} {\bibfield
  {journal} {\bibinfo  {journal} {Ann. Math. Statist.}\ }\textbf {\bibinfo
  {volume} {24}},\ \bibinfo {pages} {265} (\bibinfo {year} {1953})}\BibitemShut
  {NoStop}%
\bibitem [{\citenamefont {Raginsky}(2011)}]{Raginsky}%
  \BibitemOpen
  \bibfield  {author} {\bibinfo {author} {\bibfnamefont {M.}~\bibnamefont
  {Raginsky}},\ }in\ \href {\doibase 10.1109/ISIT.2011.6033729} {\emph
  {\bibinfo {booktitle} {2011 IEEE International Symposium on Information
  Theory Proceedings}}}\ (\bibinfo {year} {2011})\ pp.\ \bibinfo {pages}
  {1220--1224}\BibitemShut {NoStop}%
\bibitem [{\citenamefont {Guff}\ \emph {et~al.}(2019)\citenamefont {Guff},
  \citenamefont {McMahon}, \citenamefont {Sanders},\ and\ \citenamefont
  {Gilchrist}}]{GUFF}%
  \BibitemOpen
  \bibfield  {author} {\bibinfo {author} {\bibfnamefont {T.}~\bibnamefont
  {Guff}}, \bibinfo {author} {\bibfnamefont {N.~A.}\ \bibnamefont {McMahon}},
  \bibinfo {author} {\bibfnamefont {Y.~R.}\ \bibnamefont {Sanders}}, \ and\
  \bibinfo {author} {\bibfnamefont {A.}~\bibnamefont {Gilchrist}},\ }\href
  {https://arxiv.org/abs/1902.08490} {\  (\bibinfo {year} {2019})},\ \bibinfo
  {note} {eprint arXiv:1902.08490}\BibitemShut {NoStop}%
\bibitem [{\citenamefont {DeBrota}\ and\ \citenamefont
  {Stacey}(2019)}]{FAQBism}%
  \BibitemOpen
  \bibfield  {author} {\bibinfo {author} {\bibfnamefont {J.~B.}\ \bibnamefont
  {DeBrota}}\ and\ \bibinfo {author} {\bibfnamefont {B.~C.}\ \bibnamefont
  {Stacey}},\ }\href {https://arxiv.org/abs/1810.13401} {\  (\bibinfo {year}
  {2019})},\ \bibinfo {note} {eprint arXiv:1810.13401}\BibitemShut {NoStop}%
\bibitem [{Note1()}]{Note1}%
  \BibitemOpen
  \bibinfo {note} {The word `sense' may need qualification here, since the
  nature of the experience still bears the indelible mark of the equipment
  through which the atoms are `sensed'. For this kind of
  sensing-through-equipment we might borrow the word \protect \textit {Umsicht}
  from Heidegger.}\BibitemShut {Stop}%
\end{thebibliography}
\end{document}